# SOAP SERIALIZATION EFFECT ON COMMUNICATION NODES AND PROTOCOLS

Ali Baba Dauda[1], Mohammed Sani Adam[2], M. A. Mustapha[3],
Audu M. Mabu[4], Suleiman Mustafa[5]

[1,3]University of Maiduguri, Nigeria
[2]University of Malaya, Malaysia
[4]Yobe state University, Nigeria
[5]University of Iwate, Japan

**Abstract -** Although serialization improves the transmission of data through utilization of bandwidth, but its impact at the communication systems is not fully accounted. This research used Simple Object Access Protocol (SOAP) Web services to exchange serialized and normal messages via Hypertext Transfer Protocol (HTTP) and Java Messaging System (JMS). We implemented two web services as server and client endpoints and transmitted SOAP messages as payload. We analyzed the effect of unserialized and serialized messages on the computing resources based on the response time and overhead at both server and client endpoints. The analysis identified the reasons for high response time and causes for overhead. We provided some insights on the resources utilization and trade-offs when choosing messaging format or transmission protocol. This study is vital in resource management in edge computing and data centers.

## 1. INTRODUCTION

Speedy and dependable information exchange have become a necessary requirement for every communication. Information are processed and exchanged between nodes that require high speed and strong bandwidth [1]. Applications in distributed systems leverage resources at the network nodes to process messages. This leverage minimizes the dependency on the cloud servers hence reduces the latency and bandwidth.

Extensible MarkUp Language (XML) is one of the message exchange formats created to store and transport structured message [2]. XML uses Simple Object Access Protocol (SOAP) to exchange message. The SOAP is a messaging protocol for exchanging structured information, serving as an intermediate language for the applications on Internet nodes to communicate to each other. SOAP enveloped the XML document in a structured format to be sent and received across the nodes. In distributed Systems, SOAP operates based on Web Services to communicate and exchange information.

Web Services are collection of standards that combined to offer services one node to another [3]. The Web Service uses Internet technology like the Hypertext Transfer Protocol (HTTP) as a protocol to transport data [4]. Although, other protocols such as the Java Messaging System (JMS) protocol, File Transfer Protocol (FTP) and (SMTP) also exist for the exchange of services provided by Internet devices, HTTP has been the standard for such exchange [5]. Hence, the SOAP as an XML-based protocol can be exchanged through various data formats and transport protocols.

Message serialization is one of the formats of information exchange among the applications in distributed systems [6]. The serialization process translates object into stream of bytes and transmits over the network [7]. Deserialization on the other hand, converts back the streamed bytes into its original form [8]. Serialization eliminates the need for procedure calls by creating object bytes of the data and works well for processing transitory data. These attributes made serialization to improve response time in communication.

In this work, we exchanged serialized SOAP messages using two web services via HTTP and JMS protocols. Serialized and unserialized (normal) messages of sizes between 1 MB to 22 MB as payloads were exchanged using the two protocols. The fundamental effects of serialization and exchange of high payload among communication devices were studied and analyzed. The causes of such effects on Web services were discussed and possible recommendations were issued.

Rest parts of the paper are arranged as follows: Section 2 present review of related literature on serialization of SOAP, Section 3 provide the methodology of the work. Section 4 present the results, the discussions, and the key findings of the research. Section 5 provides the conclusion and direction for further studies.

## 2. RELATED WORK

Several studies on message serialization were conducted to facilitate the exchange of information across distributed systems or applications. The aim of message serialization is to reduce communication cost by improving the communication process. Relevant studies have identified or improved on the exchange of information using different serialization techniques.

A study conducted by [9] made a comparison between binary and XML serialization on .net and Java platforms. The authors used object types to compare the two data formats. The study presented a breakdown of serialization effect by analyzing duration taken by an object to be serialized in the memory on each platform. In the findings, Java proved to be better in handling binary serialization than the .net. On the other hand, .net is better in supporting deserialization of any object.

A study carried out by [10], optimized Web services performance by improving the serialization. The improvement was attained by implementing a re-use algorithm to perform a differential serialization on the message. The algorithm used an inbuilt function to compare the message structure of the previous message and an outgoing message and save the similarity as template. In this case, only unrelated part of the structure is processed and exchanged. The process minimizes the computing overhead, especially at the server. The disadvantage comes from the gradual formation of template in the entire transaction, especially if most messages failed to have similarity.

An attempt on deserialization to improve SOAP Web services performance was accomplished by [11] through deserializing incoming message. The message is then linked to some internal automata to match with a stored message for resemblance. To reduce the deserialization process, only the unrelated section of the sent message is treated. The outcome of this technique decreased the round-trip, client-side overhead and the response time. But the impediment of this technique is the lack of Garbage Collection (GC) procedure in the implementation, causing the likelihood of slowing the performance.

Study by [12] provided an insight on SOAP serialized payload exchange over HTTP. In the study, normal and serialized messages were exchanged, the memory utilization and response time were recorded at both the server and the client endpoints. The study revealed that both serialization and deserialization have the same effect on the resources and exchanging a large payload over HTTP is more cost effective than exchanging small payload. However, the normal payload has lower response time than the serialized payload. The study uses only one exchange protocol in the implementation to examine the effect of the serialization.

These studies investigated the effect of serialization and/or improved on the performance of SOAP. To complement the previous studies, our research identified the causal effects of the overheads and the response time of exchanging serialized high-volume messages via different transport protocols. The outcome can provide a new dimension of how improving SOAP performance can enhance the data exchange on distributed systems.

## 3. METHODOLOGY

We carried out the experiment using HP Desktop computer running windows 10, with internal features: Core i7, with processor speed of @ 2.30 GHz processor. The language platform is JAVA 2EE with Weblogic 12c as running server. Two implemented web services were used as SOAP via HTTP and SOAP via JMS. Two serialized and unserialized (normal) messages as payloads were exchanged between sender (server) and the receiver (client). Figure 1 demonstrates the client- server model with serializer module at the server and the deserializer module at the client.

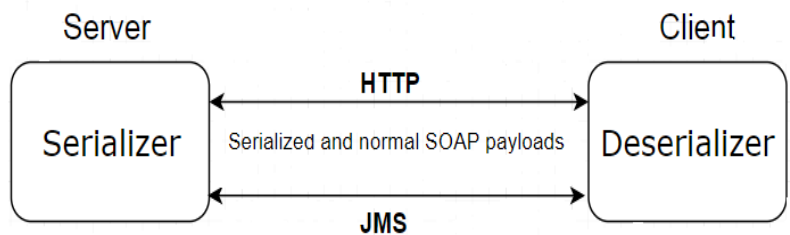

**Figure 1- Client – server model for the exchange of the normal and serialized payloads using exchange protocols**

The web services contained SOAP with HTTP binding and SOAP with JMS binding. Both web services contained classes with serializer method at the sender application (server) endpoint and deserializer method at the receiver application (client) endpoint. The server contains the function for generating payload while the client contains web service port for accessing the service from the server. The generated payload is sent from the server and received by the client. The response time and overheads at each endpoint is automatically captured and recorded. The methodology of the web services execution is shown in Figure2.

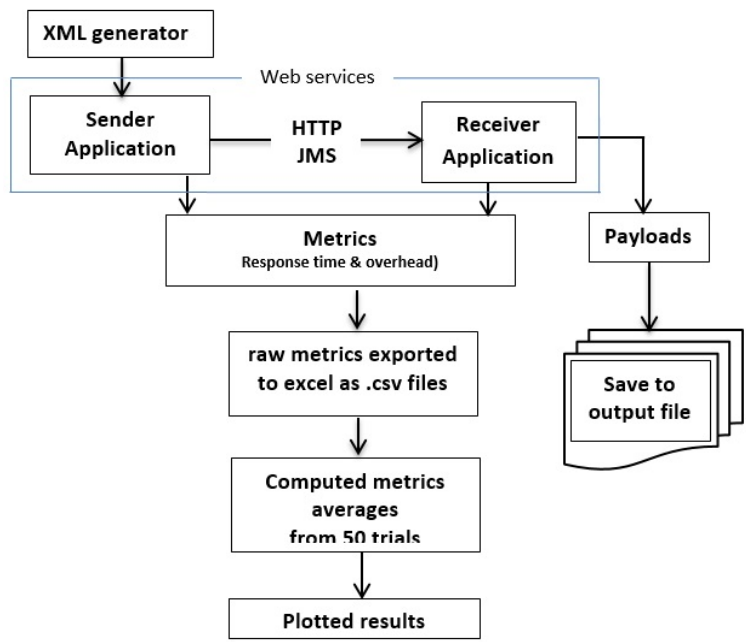

**Figure 2- Methodology flow for the web service execution**

In the JMS protocol binding, the web service is requested by the client application. The server application responds by providing the service. The server immediately creates a WSDL file containing all the information to be used by the client. The WSDL makes the service available to the client to utilize.

In the HTTP binding, the client makes the request and the server acknowledges and creates the WSDL file, and the client uses the file to find and consumes the service by providing the interface and other associated information. The server responds by directly sending the requested service using the POST method.

In this implementation, a client request for a service from the server and the server generates the request (message) and forward it to the client for consumption. The requested is processed as payload throughout the exchange process. The process is done differently for both the normal and the serialized message. The payload is added and pushed as a request to the client. The web services were executed 45 times separately and transaction metrics were recorded automatically by the applications.

## 4. RESULT AND DISCUSSION

The analysis and the discussion of the result for the two bindings: SOAP over HTTP and SOAP over JMS both the serialized and normal exchange are presented in this Section. Figure 3 – 9 depict the presentation of the findings in terms of graph.

### 4.1 NORMAL SOAP MESSAGE VIA HTTP

Figure 3 depicted the exchange of normal payload over HTTP. The transaction and client response times revealed an uneven trends with a comparable patterns as both trends ascent and descent through the exchange. The pattern continued as the payload added to 19.4 MB, and this changed the pattern of the transaction response time to sharply rose to 1551.3 ms. But the trend progressed through to the response time of 1841 ms with the ending payload of 22.2 MB.

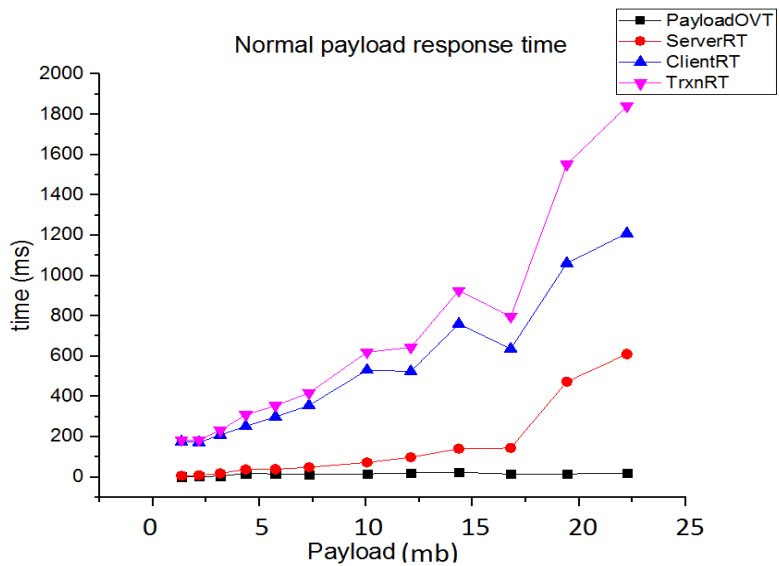

**Figure 3- SOAP over HTTP response time for normal payload transaction**

On the other hand, the server response time found to be stable in its transition with low response time throughout the transaction. Although, the transition slightly changed when the payload increased to 19.4 MB causing the response time to be 473.7 ms from 144.7 ms. However, at the final payload of 22.2 MB, the response time surged causing the transition to deviate upward. In the case of the payload generation overhead, the overhead rose gradually through the transaction causing it to formed an upward slope.

*Key findings*
The client response time revealed to be high owing to the HTTP request. In general, HTTP functions by making demand for the service from the server for every transaction. The client use the HTTP to constantly and recurrently check the server for a new message; as such, the client monopolized the exchange thread. The transaction response time indicated to be moving high with the payload increase as revealed in its trend in Figure 3, this is conceivably attributable to the client response time that took over the most part of the transaction.

The server response time rose once at the payload of 19.4MB. This might be attributed to garbage collection been invoked when the Java Virtual Machine (JVM) gathered there is insufficient free space to create new object [13]. However, this process may delay Java application for some times, resulting to hike in the server response time at that moment.

The client response time was found to be much as a result of the request by the HTTP layer. The request-based nature of the HTTP enforced the client to recheck the exchange for a new message necessary from the server, consequently, the client dominates the thread of the transaction. The transaction response time revealed to be moving upward with payload increment due to client that dominated the transaction.

The server response time was shown to be linear even though the payload is added, tends to rise slightly only at the payload of 19.4 MB. Generally, spikes and sudden drift result when the garbage collection is invoked due to insufficient memory for allocation of space to incoming messages or due to creation of new objects, as reported by [13]. The payload generation overhead revealed an undeviating trend indicated that similar processing time is required to create and link the payload to the earlier generated one.

4.2     SERIALIZED SOAP MESSAGE OVER HTTP

The serialized payload transaction over HTTP is shown in Figure 4. The metrics comprised server response time, client response time, transaction response time payload generation, serialization, and deserialization overhead.

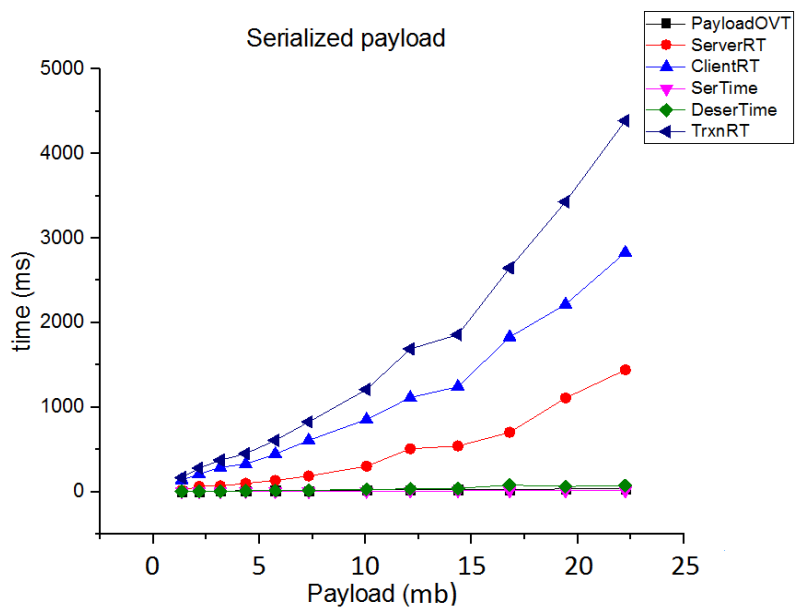

**Figure 4- SOAP over HTTP response time for serialized payload transaction**

The trend for the serialization and deserialization, and payload generation ovehead shown to be steady with almost identical delay time all through the exchange. As the exchange proceeded, there is a minor deviance in the deserialization when the payload was added to 14.4 MB, but the trend then plunged and continued the movement until the end of the exchange. The server response time is identical to the serialization and deserialization overhead at the intial stage, but went up as the payload was added to 7.3 MB. The trend raised and made more deviation up to 1442 ms with a payload of 22.2 MB.

The client and the transaction response time were discovered to be high. Although as seen in Figure 4, the two response times have similar trends at the beginning, but at the payload of 16.8 MB, each trend shot up differently and the patterns progressed matchless with different response time to the end of the transaction.

*Key Findings*

As a principle, the serialization process normally adds some fields to the total payload and the sum effect established a significant impact during the exchange, resultantly making the process to be costly. Eventually, the serialization produced performance overhead at the server. Deserialization as well, acquired overhead at the client side due to request by copying the message array and gradually assigning a fresh storage. But the linearity of the payload overhead trend implies that the same processing time is required to create and link the payload to the existing one.

The server response time is lower than the client response time, and we can make a deduction here that the frequent request and the continual check for the new message by the client cause it to dominate the transaction. Nonetheless, the reconversion of the streamed bytes to the objects form always incurs client-side overhead. The server trend rose progressively with the increase in the payload. It was observed that the payload generation overhead elevated as the payload extended to 19.4MB and gradually propagated. The possible reason might be the JVM activity, as stated in the above findings in subsection 4.1. The transaction response time exponentially drove high with the increasing payload and might be ascribed to the client response time ascendency, accounting for most of the exchange process.

4.3   NORMAL VERSUS SERIALIZED MESSAGE RESPONSE TIME FOR HTTP

The response time for SOAP payload over HTTP for both serialized and normal payload is depicted in Figure 5. The figure shows the transaction response time for each format.

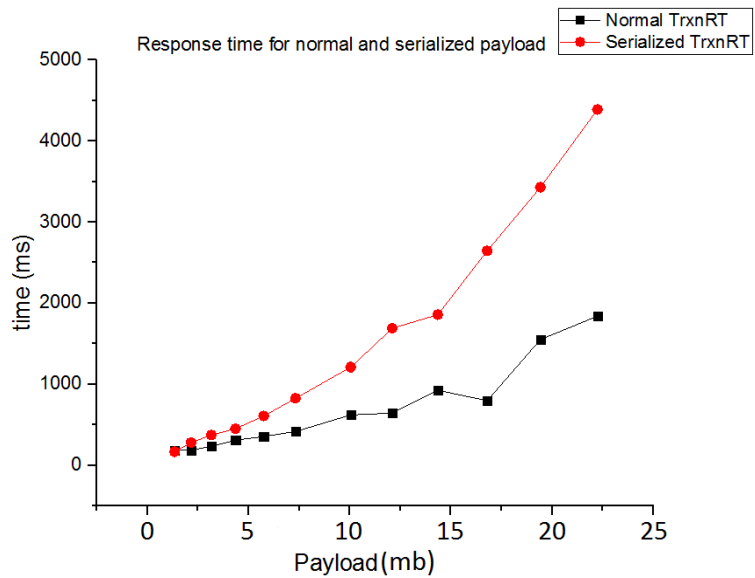

**Figure 5- SOAP over HTTP response time for serialized and normal payload**

Since the beginning, the nornal and serialized response times have no shared resemblance. The serialized payload upsurged with a rising progression as the amount of the payload is added. The trend suddenly increased when the payload surpassed 14.4 MB and the response time sped to 2646 ms. further, the trend sustained a direct upward course until the end of the exchange communication.

Though the normal payload early response time is greater than serialized payload response time, but the normal payload demonstrated a good ascension with a higher response time of 1841 ms contrasted with the serialized highest response time of 4391 ms. The response time for the serialized payload rose upright as the payload is increased and descended slightly from 925 ms to 796 ms and then rose again and moved on till the end of the communication.

The normal payload response time inclined slightly with the corresponding payload. The response time then plummeted when the payload is 16.8 MB, but then rose up and maintained the trend to the end of the transaction.

*Key Findings*
All the way through the exchange, apart from the start up point, the response time of the serialized payload is greater than the response time of the normal payload. The initial low response time by the normal payload is attributed to the possibility that the JVM made an initial caching of the payload in the memory. Figure 5, revealed that the serialized response time is incomparably higher than the normal response time. The disparity is characterized by the activity of serialization and the deserialization processes at the endpoints. The high-flown process of serialization/deserialization demands construction and reconstruction of objects, and this adds to the total overhead of the serialized payload response time. It can be concluded that in the serialized payload exchange, serialization/deserialization process contributes to most of the Central Processing Unit (CPU) utilization.

4.4    NORMAL SOAP MESSAGE VIA JMS

Figure 6 displays the server response times, client response time, the transaction response time and the payload overhead for a normal payload over JMS. As shown in the Figure 6, the response time for the server revealed to be the lowest in the exchange. The server response time started and continued slowly upright with payload increase. Although there is a slight rise in the trend when the payload increased to 19.2MB but it went down and maitained the movement.

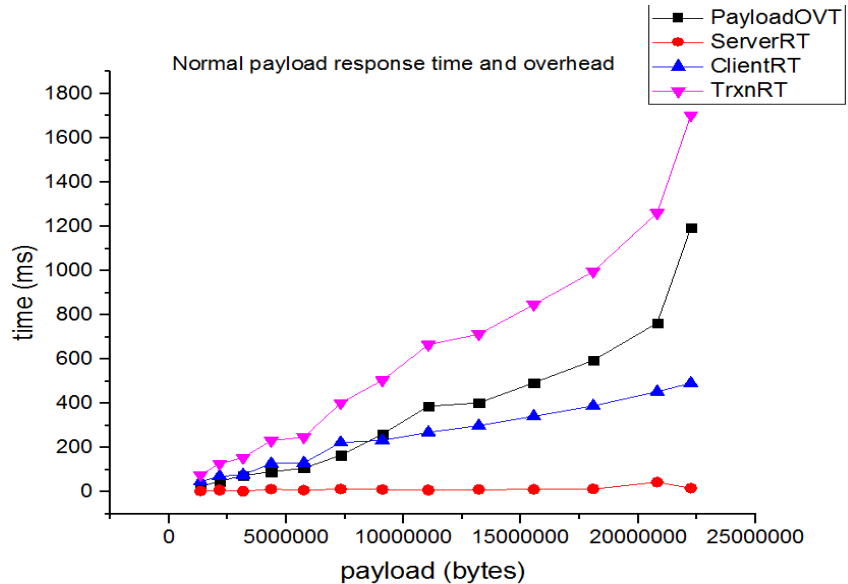

**Figure 6- SOAP over JMS for normal payload transaction comprising payload overhead, server, client and overall transaction response times**

Figure 6 revealed that the transaction response time is higher as the payload increases. The transaction response time varies accordingly at different payloads until the load was 11.9MB and caused the trend to rose slightly higher at 665.8ms. The trend then maintained the bearing linearly, but changed when the payload was 22.2MB shooting from 997.1ms to 1261.1ms.

The payload generation overhead generally rises with the increment in the payload till the end of the transaction but had an upward change when the payload increased to 11.9MB and 22.2MB. Both the transaction response time and the payload generation overhead were revealed to be high compared to other metrics.

The client response time continued to grow as the payload was increased making the transition to form a perpendicular linear progression up to the end of the transaction. In one point of a payload of 7.3MB, the response time made a sudden slight change upward to 223ms but plummeted down to maintain the course.

*Key findings*
Server response time is low because JMS is stateful and stay connected once the initial connection is established, as such little effort is required to queue the payloads. The transaction response time revealed to be much as the payload increased as shown in Figure 6. This is conceivable due to the payload overhead that occupied much part of the transaction. The payload overhead generally rises with the increment in the payload, and possibly caused by the fact that the normal payload is not restricted by any message format during the exchange process, as such the payload is not cached at the JVM. Any time the client is reading the message, it has to start over and fetches and concatenates the payload through the loop.

The client response time is higher than the server response time. This claim is evident from client response time trend line that indicates upward transition with the payload. The client starts the communication by obtaining a Java Naming and Directory Interface (JNDI) connection to the server which provides the access to the connection factory and the connection queue. The client always reads the data from queue in sequence and this process is resource demanding. On the other hand, the server produced and put the payload on the queue in any transaction. Since JMS supports different states and connects once, the server uses less resources to push the request, hence the payloads are queued with little effort.

## 4.5 SERIALIZED SOAP MESSAGE VIA JMS

The reponse time and the overhead for serialized payload over JMS is shown in Figure 7. These comprised the transaction response time, client response time, server response time, payload generation ovehead, serialization and deserialization overhead.

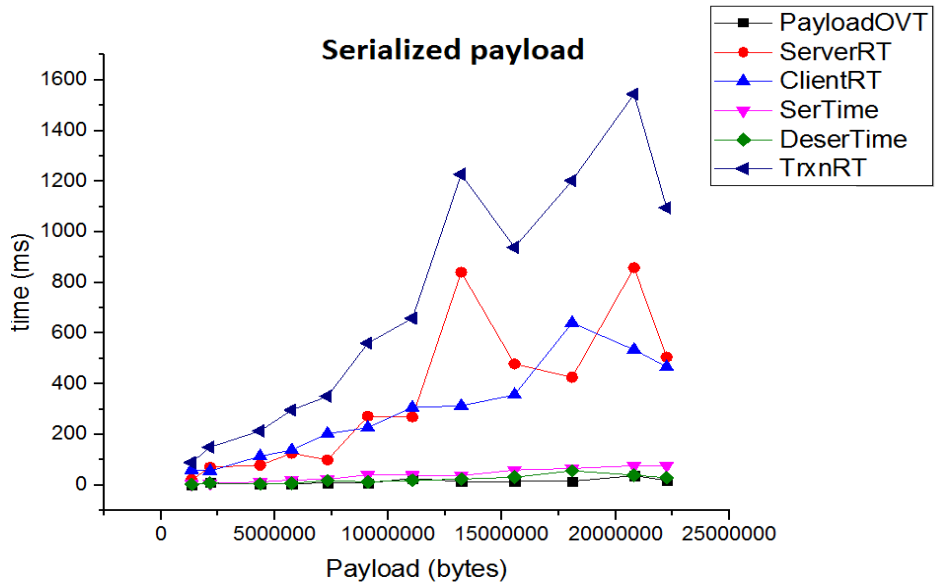

**Figure 7- SOAP over JMS metrics for the serialized payload comprising of server, client and transaction response time, and the overhead for the payload generation, serialized, deserialized**

In Figure 7, payload, serialization and deserialization overhead were marginally having similar trend through the transaction. Although the serialiazation overhead is slightly high all through, but the payload and the deserialization overhead have little variation.

The client response time was regular and moved with the growing payload between the intial payload of 1.34 MB and 15.5 MB. The trend then made an abrupt surge and ascend from 356 ms to 640 ms as the payload increased to 18 MB and gradually started to plummet despite payload being increased. On other hand, the server and the transaction response times reavealed an irregular pattern right from the inception of the transaction.

The trend of the transaction response time initial movement revealed to be normal having upward inclination from the initial workload 1.3 MB to 11 MB. The irregular movement begun when payload was at 13.2 MB with 1227 ms, but plummeted to a lower time of 939 ms despite payload increased to 15.5 MB. However, this does not stand long as the transaction time grew up to 1202 ms against 18 MB. The trend proceeded up and even higher this time with a peak time 1544 ms and payload size 20.8 MB. The load size was increased to 22 MB and the transaction response time plummeted to 1095 ms, much more lower than when the payload was 18 MB.

*Key findings*
The transaction and the client response times appeared to be high, with the transaction response time higher. The server response time maintained to be lower than the client response time since the JMS protocol connects only once, the server therefore utilizes less resources in the communication. The serialization, deserialization and the payload overheads revealed to be low with the serialization overhead slightly higher, and this generally, indicated low overhead in the SOAP over JMS exchange. The server response time grew sporadically upward forming spikes in the trend. As reported by [14], typical reasons for spike is caused by the JVM. During execution the JVM uses the garbage collection to reclaim unused memory space to allocate it to a new object and when there is no free space to be allocated. This process delays Java application for 5-25% of overall runtime [15]. In this implementation the payloads are of varying size and this forced a challenge for the JVM to handle the actual time spending on GC.

### 4.6 NORMAL VS. SERIALIZED SOAP MESSAGES RESPONSE TIME FOR JMS

The response time for serialized and normal transaction over JMS is illustrated in Figure 8. The metrics comprise transaction response time for both normal and serialized transaction.

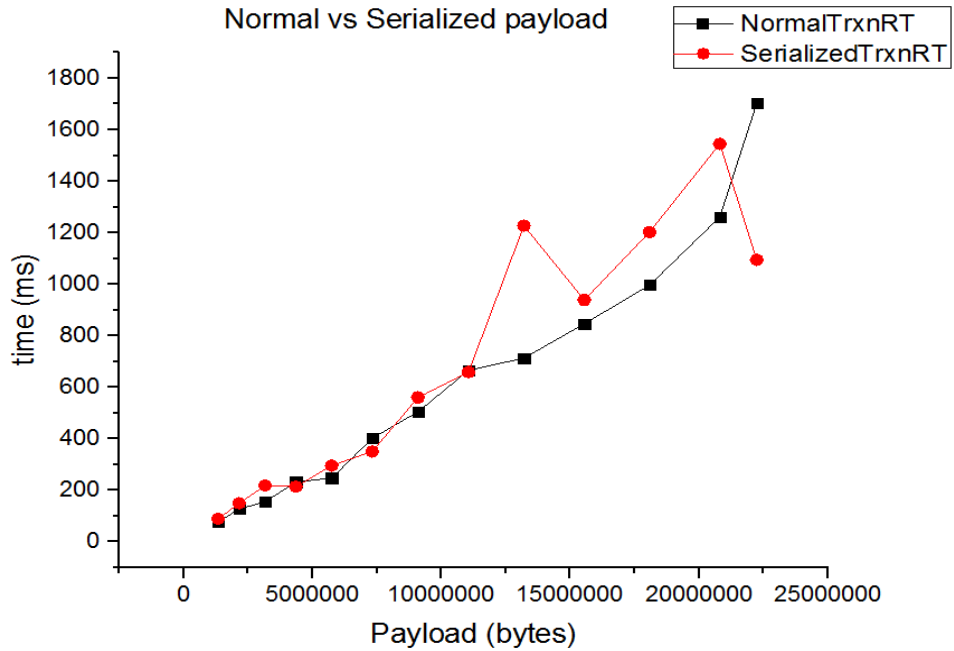

**Figure 8- SOAP over JMS response time for serialized and normal payload transactions response time**

Figure 8 revealed that both format had comparably the same pattern with almost the same response times as the trends moved in the transition until after the payload increased to 13.2 MB. In the serialized payload, the trend went up and sporadically continued the pattern, but plummeted to a lower value of of 939 ms despite payload was increased to 15.5MB though spanned for few moment as the transaction time grew up again to 1202 ms against 18 MB payload. The trend proceeded up and even higher this time with a peak time 1544 ms and payload size 20 MB. The payload size increased to 22.2 MB and the trend plummeted to 1095 ms, much more lower than when the payload was 18 MB.

The normal payload trend moved perpendicularly along the course of the transaction. This perpendicular trend continue to rise merely proportionately with the payload growth, starting as low 77 ms and smoothly risen until 1261 ms. But afterward, the trend shot up a bit as the payload was finally increased to 22.2 MB.

*Key findings*
Serialized payload response time revealed to be greater than the normal response time. The serialization and the deserialization both demand CPU resources. In JMS, the Lookup feature of the queue connection factory could not perform serialization rapidly. Also, the JMS client establishes the request and reads the messages one by one from the queue. These two resource-intensive attributes can contribute to the slow response time of the serialized SOAP over JMS. Spikes were noticed in the serialized SOAP over JMS and this is normally characterized by GC by the JVM due to CPU consumption. According to [16] this happens when the GC assumes the data object to be large hence forcing it reclaim unused memory space for a new object.

4.7     SERIALIZED AND NORMAL PAYLOADS OVER BOTH HTTP AND JMS

The transaction response times for normal and serialized payload transaction over both HTTP and JMS is depicted in Figure 9. The Figure shows the messaging format exchange using the transport protocols.

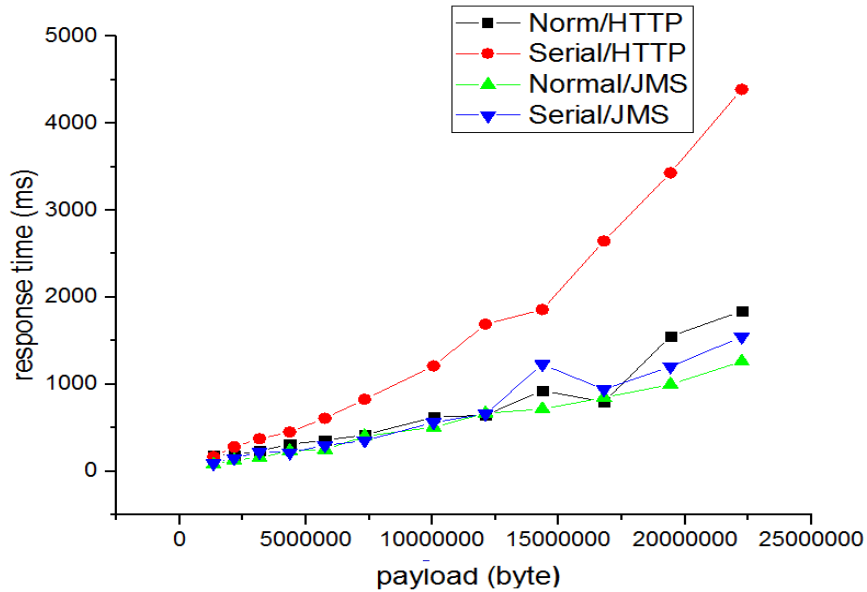

**Figure 9- Transaction response times for Serialized SOAP over HTTP, normal SOAP over HTTP, Serialized SOAP over JMS, and normal SOAP over JMS**

Serialized SOAP over HTTP is higher than the normal SOAP over HTTP, and the normal SOAP over HTTP is higher than the serialized SOAP over JMS. The serialized SOAP over JMS is higher than the normal SOAP over JMS, except in 3 points when the payload is respectively 4.4MB, 7.3MB and 12.1MB. The maximum response time for all the trend was attained by the serialized SOAP over HTTP, reaching 4391ms, while the trends for the other three trend lines were even below 2000ms. Impliedly, the normal SOAP over JMS has the lowest response time, while the serialized SOAP over HTTP has the highest response time.

The response time for the serialized SOAP over HTTP rose with the payload increase since inception and formed vertical trend. From Figure 9, it is observed that the serialized SOAP over JMS is the most resource demanding transaction.

*Key Findings*
Exchanging serialized payload SOAP over HTTP consumes a lot of resources due to two reasons; conventionally, due to HTTP request, the client always checks for incoming messages one at a time from the server. Secondly, serialization adds reference fields to the object during encoding and this reference also formed part of the serialized object as payload, thus adding to the server overhead. Spikes noticed in the serialized SOAP over JMS and the normal SOAP over HTTP might be a consequence of paging, garbage collection or memory allocation at the JVM [17]. The trend of the normal SOAP over JMS response time revealed to be the lowest. This lowest trend is caused by the JMS protocol that connects only once throughout the transaction, and moreover the normal payload is not restrained to any format.

## 5. CONCLUSION AND FUTURE WORK

In both transport protocols, serialization/deserialization consumes CPU resources, with the serialization consuming higher. Normal and serialized transaction via HTTP consumes more resources, with the serialized transaction over HTTP having the highest response time. This is possibly due to the HTTP continual request-based check for incoming message by the client. Generally, the payload overhead rises with the increment in the payload, except in the normal payload over JMS, where the payload overhead is entirely low. Possibly, this is caused by the fact that the normal payload is not constraint by any format during the exchange process as such the payload is not cached at the JVM. Considering the effects of the two binding protocols, SOAP over JMS is better than SOAP over HTTP. The response time in SOAP over JMS is low and incur less overhead. Adopting Serialized SOAP message over JMS for communication devices regardless of the volume has more benefit than the HTTP.